\documentclass[a4paper,fleqn,usenatbib]{mnras}

\usepackage{newtxtext,newtxmath}

\usepackage[T1]{fontenc}
\usepackage{ae,aecompl}

\usepackage{graphicx}	
\usepackage{amsmath}	
\usepackage{amssymb}	
\usepackage{bm}  
\usepackage{comment}
\usepackage{subfig}
\usepackage{enumitem}
\usepackage[pdftex,dvipsnames]{xcolor}
\usepackage[linecolor=MidnightBlue,backgroundcolor=MidnightBlue!25,bordercolor=MidnightBlue,prependcaption]{todonotes}

\newcommand{\eqq}[1]{Equation~(\ref{#1})}

\newcommand{\ie}{{\it i.e.}}
\newcommand{\eg}{{\it e.g.}}

\newcommand{\likeli}{\ensuremath{\mathcal{L}}}
\newcommand{\photoz}{photo-$z$}
\newcommand{\features}{\ensuremath{\mathbfit{F}}}
\newcommand{\redshifts}{\ensuremath{\mathbfit{z}}}
\newcommand{\types}{\ensuremath{\mathbfit{t}}} 
\newcommand{\positions}{\ensuremath{\bm{\mathit{\theta}}}}
\newcommand{\fields}{\ensuremath{\bm{\mathit{\delta}}}}
\newcommand{\biases}{\ensuremath{\mathbfit{b}}}
\newcommand{\fractions}{\ensuremath{\mathbfit{f}}}
\newcommand{\counts}{\ensuremath{\mathbfit{N}}}
\newcommand{\priorcounts}{\ensuremath{\mathbfit{M}}}
\newcommand{\nzt}{\ensuremath{{N}_{zt}}}
\newcommand{\mzt}{\ensuremath{{M}_{zt}}}

\title[Redshifts with colors and clustering]{Redshift inference from the combination of galaxy colors and clustering in a hierarchical Bayesian model}

\author[S\'anchez \& Bernstein]{
Carles S\'anchez\thanks{E-mail: carless@physics.upenn.edu}
and Gary M. Bernstein\thanks{E-mail: garyb@physics.upenn.edu}
\vspace{1mm}
\\
Department of Physics \& Astronomy, University of Pennsylvania, 209 S. 33rd St., Philadelphia,
PA 19104, USA
}

\date{\today}

\pubyear{2018}

\begin{document}
\label{firstpage}
\pagerange{\pageref{firstpage}--\pageref{lastpage}}
\maketitle

\begin{abstract}
Powerful current and future cosmological constraints using high precision measurements of the large-scale structure of galaxies and its weak gravitational lensing effects rely on accurate characterization of the redshift distributions of the galaxy samples using only broadband imaging.
We present a framework for constraining both the
redshift probability distributions of galaxy populations and the
redshifts of their  individual members. We use a hierarchical Bayesian model (HBM) which provides full posterior distributions on those redshift probability distributions, and, for the first time, we show how to combine survey photometry of single galaxies and the information contained in the galaxy clustering against a well-characterized tracer population in a robust way.  One critical approximation turns the HBM into a system amenable to efficient Gibbs sampling.  We show that in the absence of photometric information, this method reduces to commonly used clustering redshift estimators.
Using a simple model system, we show how the incorporation of clustering information with \photoz's tightens redshift posteriors, and can overcome biases or gaps in the coverage of a spectroscopic prior. The method enables the full propagation of redshift uncertainties into cosmological analyses, and uses all the information at hand to reduce those uncertainties and associated potential biases.  
\end{abstract}




\section{Introduction}
\label{sec:intro}

Large galaxy surveys constitute a powerful probe for testing cosmological models through the information they provide about the large-scale structure of the Universe. There exist two main categories of surveys. Spectroscopic surveys such as 
\textit{2dF} \citep{Colless2001}, 
the \textit{VIMOS-VLT Deep Survey} \citep{LeFevre2005}, 
\textit{WiggleZ} \citep{Drinkwater2010}, 
\textit{Baryon Oscillation Spectroscopic Survey} \citep{Dawson2013} or 
\textit{Dark Energy Spectroscopic Instrument} \citep{DESI2016} supply three-dimensional information about the galaxy distribution, but they are expensive in time and resources. Alternatively, imaging or photometric surveys like 
the \textit{Sloan Digital Sky Survey} \citep{York:2000gk}, 
\textit{PanSTARRS} \citep{Kaiser2000}, 
the \textit{Kilo-Degree Survey} \citep[\textit{KiDS},][]{Jong2013}, 
the \textit{Dark Energy Survey} \citep[\textit{DES},][]{Flaugher2015}, 
the \textit{Hyper-Suprime-Cam} survey \citep[\textit{HSC},][]{HSC2012}, 
or the \textit{Large Synoptic Survey Telescope} \citep[\textit{LSST},][]{LSST2012}
are more efficient per galaxy, and enable weak gravitational lensing measurements via galaxy shapes --- but do not provide a complete 3D picture of the visible Universe, as redshifts are estimated using only broadband spectra of galaxies. 

A critical task for cosmological inference using imaging surveys is to
establish the redshift distribution $n(z)=\mathrm{d}N/\mathrm{d}z\,\mathrm{d}A$ of a sample of
sources passing some selection criterion $s$, since errors in the characterization of such distributions can lead to biases in the cosmological parameter estimation \citep{Huterer2006,Hildebrandt2012,Cunha2012,Benjamin2013,Huterer2013,Bonnett2015,Hoyle2017}.  In some cases the
posterior distributions $p(z_i)$ of the redshifts of individual
sources are of use as well.
Two forms of information commonly form the basis of these inferences.
On one hand, \emph{photometric} redshifts compare a set of observed fluxes (or
colors or other measurable features) $F_i$ 
of source $i$ to those expected for galaxies at various
redshifts to infer the redshift of the individual target galaxy from different techniques, broadly classified as template fitting methods~(e.g. {\tt Hyperz}, \cite{Bolzonella2000}; {\tt BPZ}, \cite{Benitez2000,Coe2006}; {\tt LePhare}, \cite{Arnouts2002,Ilbert2006}; {\tt EAZY}, \cite{Brammer2008}), and training methods ~(e.g. {\tt ANNz}, \cite{Collister2004}; {\tt ArborZ}, \citet{Gerdes2010}; {\tt TPZ}, \citet{CarrascoKind2013a}; {\tt SkyNet}, \citet{Bonnett}). Comparisons of different methods have been tried in both simulated and real data
\citep{Hildebrandt2010,Dahlen2013,Sanchez2014}. On the other hand, with the advent of large-area surveys it has become
practical to constrain $n(z)$ using \emph{clustering} information
based on the observed sky positions $\theta_i$ of the sources. For the latter, most
applications are based on 2-point statistics of the source population
with reference to the positions of a tracer population of
galaxies having secure \textit{a priori} redshift assignments
\citep{Newman2008,Menard2013,Schmidt2013}. In addition, there is typically prior information 
on the redshift distribution of the population coming from a subset of galaxies in the survey, 
spanning a small area of the sky, for which \emph{spectroscopic} or high-quality 
photometric redshifts are available. Attempts to use both photometric and
clustering constraints on the same survey data in the presence of prior information have been made by the
\textit{KiDS} \citep{Hildebrandt2017} and \textit{DES} \citep{Hoyle2017,Gatti2018,Davis2017,DESCollaboration2017} surveys.  The two methods have been used
together either by means of basic visual cross-checks on the two independently
derived $n(z)$'s, or to provide cross-checks and joint constraints on
some single summary statistic of $n(z)$, such as its mean.

In this work we develop and test a method to produce a single
inference for $n(z)$ (and the $z_i$) that intrinsically combines prior, 
photometric, and clustering information, and includes the capability to
produce samples of $n(z)$ drawn from the posterior
distribution constrained by all the sources of information. This method is
robust to some of the practical 
difficulties of clustering or photometric redshifts, such as the
absence of tracers over some portion of the full angular or spectral extent of the
survey, or to variation of the selection function or measurement noise
across the survey footprint.  This is achieved by extending the
Bayesian hierarchical method of \citet[][LMP]{Leistedt2016} to include the
presence of fluctuations $\delta_z(\theta)$ in the density of sources at redshift $z$ and
sky position $\theta.$ Other 
changes with respect to the LMP method include the addition of an informative prior, 
and the separation of the photometric noise part of the problem from the characterization 
of the so-called \emph{color-redshift} relation, which we consider independently. 
We will demonstrate that our method reduces to the results of the
standard 2-point clustering method in the limit where the prior and the $F_i$
are uninformative and the clustering is weak.

The combination of clustering, photometric, and spectroscopic information into a single $n(z)$ inference is advantageous for two main reasons.   First, it has not been clear to date how one should combine or even check consistency between independent inferences, since each is predicting a function $n(z).$  Second, each method on its own can have failures that render it impossible to generate a usable inference: for example, clustering measures can lack a tracer population at some $z$ ranges; spectroscopic surveys are usually highly incomplete in some way which is biased with redshift; and photometric inferences are often biased by errors in assumed ``truth'' colors.  In such circumstances an inference based on a single technique can be so uncertain as to be worthless for precision cosmology.  But a combined inference could allow each method to resolve the degeneracies or ameliorate the intrinsic biases of another, and be the only route to a viable estimate.

In Section~\ref{sec:method} we derive the posterior probability for
redshift information when conditioned on observed galaxy features and
positions.  Section~\ref{sec:density} derives the optimal kernel
density estimator to be applied to a tracer galaxy population.  
In Section~\ref{sec:sampling} we outline a Gibbs sampler for the system, and a
straightforward method for inclusion of prior constraints on $n(z)$
derived from spectroscopic samples.  We then
implement the method on a simple simulation in Section~\ref{sec:sims},
demonstrating the ability of the combined inference to robustly reduce noise
and biases present in the spectroscopic or photometric information
alone.  Appendix~\ref{sec:equivalence} shows that our method
reproduces the results of standard clustering-$z$ techniques in the
absence of photometric information (under some applicable approximations).

\section{Probabilities for redshift distributions}
\label{sec:method}

Our approach closely parallels that of LMP, with the colors of individual galaxies $i$ are seen as being drawn from a pool of possible types $t_i,$ redshifts $z_i,$ and angular positions $\theta_i$ with some latent intrinsic mean density $n(t,z)$ on the sky, with observations then yielding a noisy set of observable features which we denote as $F_i$. The critical extension to the LMP model is to consider that the galaxy densities at redshift $z$ vary by some factor $1+\delta_z(\theta)$.  The tracer population used for correlation redshifts is drawn from the same latent density distribution (up to some bias factor).  Note also an important difference from the LMP model and notation, whereby we define galaxy ``types'' by observed properties rather than rest-frame properties. This \emph{phenotype} approach will allow us to decouple the measurement-noise part of the problem from the color-redshift relation. Our notation will be that the vector quantities \features, \types, \redshifts, and \positions\ denote the full set of properties of all selected galaxies, \ie\ $\features=\{F_1,F_2, \ldots, F_N\}$ (a summary of all the notation can be found in Table \ref{tab:notation}).  

\subsection{Generative model}

Our fundamental assumption will be that the galaxies are drawn from a 
\textit{Cox process} \citep{Cox1955}
or \textit{doubly stochastic Poisson process}, whereby a point set is
Poisson sampled from a latent, stochastic density field.  We simplify
the problem to consider that the redshift $z$ is transformed to an
integer indexing a set of finite-width redshift bins, and each bin has
an independent
density fluctuation field $\delta_z(\theta)$, with $\langle
\delta_{z_i}(\theta) \delta_{z_j}(\theta)\rangle = 0$ for $z_i\ne
z_j.$  

We first consider a simple case in which the sky is populated with galaxies that have identical appearance on the sky (\ie\ constant $F$) and
a mean density per unit solid angle of $n.$
The galaxies have intrinsic redshift distribution specified by
$f_z=p(z),$ with $\sum_z f_z=1.$ The galaxies' spatial distribution has linear bias $b_z$ with respect to
the underlying density fluctuation $\delta_z$.  We also assume there is some selection function $s$ with the probability of a galaxy being selected perhaps depending on sky position, specified as a selection or window function $p(s|\theta).$ We will always assume that we know nothing about the non-selected galaxies, not even that they exist; the observed data are the positions \positions\ and features \features\ of the selected sources. The selected galaxies can be considered now as being a Poisson sampling of a density field
\begin{equation}
  \rho(z,\theta | n, \fractions, \biases, \fields) = n f_z \left[1+b_z\delta_z(\theta)\right] p(s|\theta).
  \label{density1}
\end{equation}
Note that the last two terms describe the spatial variation of expected detection rate due to density fluctuations and variable observing conditions, respectively.
Then the probability of selecting a set of galaxies $i\in \{1 \ldots N\}$ at positions \positions\
and redshifts \redshifts\ takes the standard Poisson form:
\begin{align}
  p(\redshifts,\positions | n, \fractions, \biases, \fields) &
	= \exp\left\{-\int \mathrm{d}^2\theta \sum_z \rho(z,\theta)\right\} \prod_{i=1}^N \rho(z_i,\theta_i)
    \nonumber  \\
 &  = \exp\left\{-n \int
  \mathrm{d}^2\theta\, p(s | \theta) \sum_z f_z \left[1+b_z\delta_z(\theta)\right]
\right\} \nonumber \\
 & \phantom{=} \times 
\prod_{i=1}^{N} p(s | \theta_i)\times n\, f_{z_i}
\left[1+b_{z_i}\delta_{z_i}(\theta_i)\right].
\label{pzt1}
\end{align}
Note that the normalization part of the Poisson likelihood in the exponent contains a sum over $z$ rather than an integral, because we have made $z$ a discrete variable.

\begin{table}
	  \caption{Summary of the notation used throughout this paper.}
    \label{tab:notation}
    \begin{center}
    \begin{tabular}{cl} 
      \hline
	    $F$ & galaxy set of observed features \\
	    $t$ & galaxy phenotype (or simply type) \\
	    $z$ & galaxy redshift \\
            $\theta$ & galaxy angular position in the sky \\
            $s$ & indicator of successful detection/selection \\
	    $\likeli_{it}$ & probability of measuring galaxy $i$ with $F_i$ given $t$ \\
	    $\features, \types, \redshifts, \positions$ & set of properties for all galaxies in the sample  \\
	    $N$ & number of galaxies in the sample \\
	    $N_t$ & number of types \\
	    $N_z$ & number of redshifts \\
	    $A$ & effective area of the survey for source detection \\
	    $n$ & mean galaxy density per unit solid angle \\
	    $n(z)$ & mean galaxy density per unit solid angle per $z$ \\
	    $\delta_z(\theta)$ & density fluctuation at a given $z$ and $\theta$ \\
	    $\pi_\delta$ & density fluctuation field hyperparameters \\
	    $\fields$ & set of $\delta_z(\theta)$ for all redshifts and  positions \\
	    $b_z^t$ & linear galaxy bias for type $t$ at redshift $z$ \\
	    $\biases$ & set of $b_z^t$ for all types and redshifts \\
	    $f_{zt}$ & joint type and redshift probability $p(z,t)$  \\
	    $\fractions$ & set of $f_{zt}$ for all types and redshifts \\
	    $\nzt$ & number of sources assigned to redshift $z$ and type $t$ \\
	    $\counts$ & set of $\nzt$ for all redshifts and types \\
	    $\mzt$ & number of sources in the prior at redshift $z$ and type $t$ \\
	    $\priorcounts$ & set of $\mzt$ for all redshifts and types \\
	    $\Delta z$ & difference between the means of \\ 
	     & estimated and true $n(z)$'s \\
      \hline
    \end{tabular}
    \end{center}
\end{table}

Next we generalize \eqq{pzt1} to allow for variety among galaxies.  We
define a \textit{phenotype} of galaxy to specify its noiseless,
observed appearance.  In other words, all galaxies of phenotype $t$ observed in the same conditions
are assumed to have the same selection function $p(s | t, \theta)$ and same probability
$p(F, s | t, \theta)$ of  being selected and measured to have image
features $F$.  
Following biological nomenclature, the phenotype strictly describes
the manifestation of the galaxy in the image.  Galaxies of identical phenotype
$t$ could live at distinct redshifts.
The typical
astronomical definition of galaxy ``type'' specifies rest-frame
properties, but a fixed type may have observed properties that are
redshift dependent.   Our ``phenotype'' reverses these, and will allow
us to use well-observed sources as a library of phenotypes even when
knowledge of these sources' redshifts is imprecise. Such galaxy type 
definition allows us to decouple the measurement-noise part of the problem
from the color-redshift relation.  The latter will be informed by prior spectroscopic observations or models, 
and by the clustering information. 

We will assume that we have a finite set of phenotypes indexed by integer $t$.
Each has a mean sky density of $n^t = n f_t$ where we place $n$ as the total density of all detectable galaxy phenotypes, and $f_t = p(t)$ being the fraction of the population in each type.  We have the constraint $\sum_t f_t=1.$  
We write $p(z | t)=f^t_z$ for the redshift distribution of type $t$, and we will also denote
\begin{equation}
	\label{pzt}
f_{tz} \equiv p(z,t) =p(z | t) p(t) = f^t_z f_t.
\end{equation}
Letting the bias also depend on phenotype as $b^t_z$, we can generalize \eqq{density1} to the expected mean density for each phenotype $t$:
\begin{equation}
  \rho(z,\theta, t | n, \fractions, \biases, \fields) = n f_{tz} \left[1+b_z^t\delta_z(\theta)\right] p(s|t, \theta).
  \label{density2}
\end{equation}
Knowing the survey noise properties and the intrinsic (noiseless) appearance of phenotype $t$, we can also determine the likelihood $p(F, s | t, \theta, z)$ of a galaxy of phenotype $t$ at location $\theta,z$ being selected and measured with features $F$.  This function will be independent of $z$ since the phenotype's observables are, by definition, independent of $z$.  Thus for each observed galaxy $i$ and phenotype $t$ we can assign a feature/selection likelihood 
\begin{equation}
  \likeli_{it} \equiv  p(F_i, s | t_i, \theta_i).
\end{equation}
This function depends on the quality of the observations in direction $\theta_i$ and the selection/measurement algorithms. We will assume that this is known \textit{a priori}, \eg\ by the result of injecting artificial copies of the phenotype into the images \citep{Suchyta2016}.

We can now extend the Poisson probability (\ref{pzt1}) to include a variety of phenotypes and the probability observing the galaxy features:
\begin{align}
\label{full1}
p\left(\features, \positions, \types, \redshifts | n, \fractions,\biases, \fields\right)
  & = \exp\left[-n \sum_t f_t A^t(\fractions^t, \biases^t, \fields)\right] \\ 
  & \phantom{=} \times 
                    \prod_i \likeli_{it} n f_{t_i z_i} \left[1+ b^{t_i}_{z_i} \delta_{z_i}(\theta_i)\right].
\nonumber 
\end{align}
The exponentiated quantity is, as required for Poisson distributions, the expected number of detections $\langle N \rangle$ for the entire sample.  We express this in terms of an effective survey area for each phenotype $t$:
\begin{align}
\nonumber
A^t(\fractions,\biases,\fields) & \equiv
      \sum_z  \int \mathrm{d}^2\theta\, p(s | t, \theta) f^t_z\left[1 + b^t_z
                        \delta_z(\theta)\right] \\
 & = \int \mathrm{d}^2\theta\, p(s | t, \theta) \left[1 + \sum_z   f^t_z b^t_z
                        \delta_z(\theta)\right] \\
 & \approx \int \mathrm{d}^2\theta\, p(s | t, \theta)
\label{aeff}
\end{align}
which can also be determined from knowledge of the survey properties.  In the last line we make the approximation that the density fluctuation integrated over the mask is small, $\int \mathrm{d}^2\theta \, p(s|\theta) \delta_z(\theta)=0.$ In this case the $A^t$ are known constants of the survey.

\begin{figure}
\includegraphics[width=60mm]{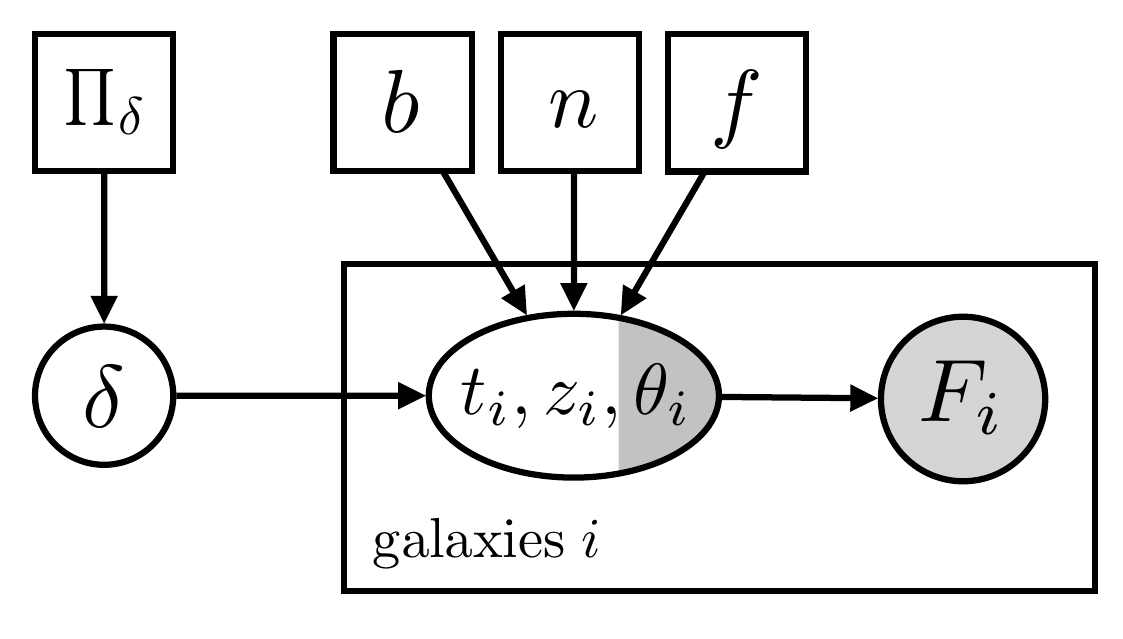}\centering
\caption[]{A directed acyclic graph of the hierarchical model for the observable positions \positions\ and features \features\ in the galaxy catalog.  Observed quantities are shaded: only the positions $\theta_i$ and image features (\eg\ fluxes) $F_i$ of individual sources, not types or redshifts. Squares hold model parameters, circles are stochastic quantities.}
\label{dag}
\end{figure}

The full generative model for the data must also specify
the process $p(\fields | \pi_\delta)$ generating the stochastic density
fluctuation fields given hyperparameters $\pi_\delta,$---\eg\
a log-normal distribution with $\pi_\delta$ specifying the power spectrum.  We also require priors $p(\biases)$ and $p(n)$, plus any prior information on $p(\fractions)$ aside from the constraint that $\sum_{tz} f_{tz}=1.$  Figure~\ref{dag} is a directed acyclic graph of the model for generating the galaxy data.

\subsection{Redshift inference}
The principal quantity of interest is the underlying redshift distribution
\begin{equation}
n(z) = n \sum_t f_{tz}.
\end{equation} 
The equation can easily be altered to yield the redshift distribution of chosen subsets of the phenotypes. We will usually be concerned with the shape, not the normalization, of $n(z)$, so will pay little attention to $n$ and focus on the fractions \fractions.   We will also be interested in many cases in the probabilities of the individual redshifts \redshifts, and to enable a Gibbs sampling scheme we will keep \biases\ and \types\ as conditional variables.  Using Bayes' theorem, the posterior joint probability of these variables of interest is
\begin{align}
\label{pfz1}
p(\fractions, \redshifts, \biases, \types | \features, \positions, \pi_\delta) & \propto
 \int \mathrm{d}n \,  \mathrm{d}\fields  \\
  & \phantom{=\int} 
  p\left(\features, \positions, \types, \redshifts | n, \fractions, \biases, \fields\right)  p(\fields | \pi_\delta)  \, p(n) \, p(\fractions) \, p(\biases).
\nonumber
\end{align}

The first term under the integral is given by \eqq{full1}.  We take the standard scale-free Jeffreys (logarithmic) prior $p(n)$ on the overall density $n$. Marginalization over $n$ then yields:
\begin{align}
\label{pfz2}
p(\fractions, \redshifts, \types, \biases | \features, \positions, \pi_\delta) & \propto
p(\fractions) \, A(\fractions)^{-N}\, p(\biases) \int \mathrm{d}\fields \,  p(\fields,\pi_\delta) \\
& \phantom{\propto}
\times \prod_i \likeli_{it} f_{t_iz_i}
\left[1+ b^{t_i}_{z_i} \delta_{z_i}(\theta_i)\right] \nonumber \\
A(\fractions) & \equiv \sum_{zt} f_{tz} A^t.
\end{align}
The total effective area of the survey for source detection is $A(\fractions)$, such that $\langle N \rangle = nA(\fractions).$

The required marginalization of the Cox distribution over the stochastic
field(s) in \eqq{pfz2} is known to be analytically intractable for all but the most
trivial processes $p(\fields | \pi_\delta)$.  For some processes
(including log-normal) there are feasible means to sample over the
joint distribution of $\fields$ and the other parameters, at which
point the marginalization is trivial---an example of log-normal field
inference from \photoz\ data is \citet{Jasche2012}.  A full treatment of the Cox posterior would enable the clustering redshift inference to make use not only of correlations between a tracer population and the target population, but also the clustering among target populations. We will defer any
attempt at sampling $\fields$ to further work. 

In this paper we will instead work with the approximation that \emph{we can
  replace the stochastic density fluctuation $\delta_z(\theta)$ with
  some determinsitc estimator $\hat\delta_z(\theta)$ of the realization of the density fields}  in the generative
probability of \eqq{full1}. In other words we will replace the $p(\fields | \pi_\delta)$ in \eqq{pfz2} with a Dirac delta function enforcing $\delta_z = \hat \delta_z$, which also renders the 
hyperparameters $\pi_\delta$ as irrelevant. The estimator $\hat\delta_z$ will presumably come from a tracer population; this approximation loses any information that may be present in the clustering amongst the target population.  We will refer to this as the ``deterministic density'' approximation, since it yields a single set of density fields $\hat\delta_z(\theta)$ from the observed set of tracers, ignoring the fact that the density fields remain stochastic even after the tracer population is specified.
With this simplifying approximation, the posterior distribution for redshift information in \eqq{pfz2} becomes
\begin{align}
\label{pzf3}
p(\fractions, \redshifts, \types, \biases | \features, \positions) & \propto
p(\fractions) \, A(\fractions)^{-N} 
 \prod_i  \likeli_{it_i} \, f_{t_iz_i} \, \left(1+ b^{t_i}_{z_i} \hat\delta_{iz_i}\right), \\
\hat\delta_{iz} & \equiv \hat\delta_z(\theta_i). \end{align}

The roles of the photometric and clustering information are clear and simple in the posterior of \eqq{pzf3}.  The photometric information is in $\likeli_{it}$, which tells how well galaxy $i$ resembles phenotype $t$ (along with how likely this is to pass selection).  We may, for instance, know that our feature vector is equal to some ``truth'' value $F_t$ associated with phenotype $t$, with some additive noise determined by the observing conditions in direction $\theta_i$ of the survey: $\likeli_{it} = \likeli(F_i - F_t | \theta_i).$  Next is the term $f_{tz},$ which is an expression of the prior probability that any galaxy is of phenotype $t$ and redshift $z$.  Finally there is the clustering term, describing the modification of the probability by (our estimator for) the density fluctuation field.  

The technique for combined photometric/clustering redshift inference is thus as follows:
\begin{enumerate}
\item Create a library of phenotypes $t$, which span the range of possible feature vectors and window functions that galaxies can exhibit in the survey.  The phenotype library might be developed, for instance, from the galaxies detected in a deep subset of the imaging survey.  One possible approach is to discretize the color space of galaxies using a self-organized map \citep{Masters2015}, in which case the cells of such map would become our galaxy phenotypes, representing well localized regions of observed feature space.
\item Characterize the window functions $p(s | t, \theta)$.  If the selection function is well chosen, then this may be analytically accessible---for example if $s$ is a sharp flux cutoff, and the flux measurement is known to have Gaussian errors, then the selection becomes an error function.  Alternatively one can inject copies of $t$ into the data and empirically determine the selection rate, as in \citet{Suchyta2016}.  Then we integrate $p(s|t, \theta)$ over the survey footprint to get the effective areas $A^t$ for each type, and evaluate the $\likeli_{it} = \likeli(F_i, s | t, \theta_i)$ for each detection.
\item Using the tracer population, develop an overdensity estimator $\hat\delta_{iz}$ at each redshift $z$ and the sky position $\theta_i$ of each detection.  
\item Posit a prior $p(\biases)$ for the clustering biases $b^t_z$ of each phenotype relative to the tracer population.  In practice, this prior will need to bound the possible dependence of bias on redshift and type if we want to avoid the bias-density degeneracy inherent to clustering information.
\item Posit a prior $p(\fractions)$ for the fractions of the galaxy population in each type/redshift combination.  This prior would likely limit the $z$'s available for a given $t$ based on astrophysical knowledge that the colors/features of phenotype $t$ are consistent with a small number of possible redshifts, and/or by knowledge of the redshifts obtained from spectroscopy of some small sample from this phenotype.
\item Sample from the posterior distribution of \fractions, \redshifts, \types, and \biases\ defined by \eqq{pzf3}.  From these samples one can construct samples from the posterior distributions of $n(z)$ and/or the individual redshifts $z_i$ as marginalized over type and bias.
\end{enumerate}

One important caveat to the replacement of the Cox process with a deterministic density estimator is that the samples of posterior redshift information that we extract from this process will not include the variance attributable to noise in the density estimator, \ie\ the shot noise in the tracer population.

\section{Density estimators}
\label{sec:density}
Clustering redshifts make use of a population of galaxies with reliably known redshifts, either from spectra or from highly accurate \photoz's.  The tracer population must have a known selection function in a given redshift bin so that meaningful clustering statistics can be produced.  
The proper way to use the tracer galaxies would be to
define a phenotype $t=T\!z$ for the tracers in each redshift bin $z$, and include them in the full Cox process sampling.  For the tracer phenotypes, we would have
$f^{T\!z}_{z^\prime}=\delta_K(z,z^\prime)$, and could set $b^{T\!z}_z = 1,$
normalizing the $\delta_z$ to the fluctuations of the tracers.  

But as noted we will instead pull the tracer galaxies out of the likelihood and use them to produce a deterministic density estimator $\hat\delta_z(\theta).$
We can also define our estimator to yield
$\langle \hat \delta_z \rangle=0$ when averaged over the survey footprint.  

\subsection{Optimal kernel density estimator}

We investigate a kernel density estimator (KDE) for
$\hat\delta_z(\theta).$ We will show here that this
yields posterior likelihoods based on pair
counts.  Standard methods for correlation redshifts estimate $n(z)$
using two-point correlation functions, which are also computed from
pair counts, but the standard methods never attempt to make point
estimates of the fluctuation fields.  We will show here and in
Appendix~\ref{sec:equivalence} that, in a case where photometric
information (\features) is ignored, the use of an optimal KDE in our
method yields $n(z)$ estimates that are equivalent to the standard
two-point methodologies.

In this section we will drop the subscript $z$ on the assumption that we will
treat distinct redshift bins as having independent mass
distributions.  With the assumption that the kernel depends only on
the distance $\theta_{xT}$ between a test point $x$ and a tracer
galaxy $T,$ we have
\begin{equation}
\label{kde1}
\hat\delta(x) = \frac{1}{n_T}\sum_T K(\theta_{xT}) - \frac{1}{n_R}\sum_R K(\theta_{xR})
\end{equation}
with $n_T$ the mean density of tracers, and $n_R$ is the mean density
of a population of unclustered random points having the angular selection
function of $T$. We assume $n_R\gg n_T$ so that the latter term can be
considered a deterministic integral over survey area.
We assume that the process for $\delta$ is stationary and isotropic,
with known variance $\sigma^2 = \langle \delta^2 \rangle$ and
correlation function $w(\theta_{12})=\langle \delta(\theta_1)\delta(\theta_2)\rangle.$ These conditions are easily met when using cosmological tracer populations.

We seek a kernel that minimizes the error $\left\langle
  (\hat\delta-\delta)^2\right\rangle$ while maintaining no bias, 
$\left\langle \hat\delta | \delta \right\rangle = \delta$ for any
field point.  We can construct this minimum-variance estimator
straightforwardly under two conditions:
\begin{itemize}
\item The field has a function $r(\theta)$ such that $\left\langle
    \delta(\theta_1) | \delta(\theta_2) \right\rangle =
  r(\theta_{12})\delta(\theta_2).$ It is straightforward to show that
  if there is a function $r$ that satisfies this equation, then it must be
 $r(\theta)=w(\theta)/\sigma^2=w(\theta)/w(0)$.  For a Gaussian field, the mean $\delta(\theta_1)$
 conditioned on $\delta(\theta_2)$ is indeed linear in $\delta(\theta_2).$
\item The noise in the KDE estimate is dominated by Poisson shot
  noise.  This holds in the limits of low tracer density or weak
  clustering.  More generally the variance of correlation estimators
  involves sample-variance terms \citep{Bernstein1994}, but we will ignore this.
\end{itemize}
In the real cosmological applications, the first condition is not necessarily true because the density field is non-Gaussian at the smaller ($\sim10$~Mpc) scales where most of the kernel's power will be.  The Poisson-dominated limit will typically be attained for tracer populations at the relevant scales.  Should either assumption fail, it merely means that the KDE we derive here could be non-optimal; it does not invalidate the approach.  Future work will examine these effects with simulations of the real sky.

When these assumptions are true, the ideal kernel will minimize
\begin{equation}
{\rm Var}\left[\hat\delta(x)\right] = \frac{1}{n_T} \int 2\pi\theta\,
\mathrm{d}\theta\, K^2(\theta)
\end{equation}
subject to the constraint 
\begin{equation}
\int 2 \pi \theta\, \mathrm{d}\theta\, K(\theta) r(\theta) = 1
\end{equation}
which is optimized by
\begin{equation}
\label{kde2}
K(\theta) = \frac{r(\theta)}{\int 2\pi \theta^\prime\,
  \mathrm{d}\theta^\prime\, r^2(\theta^\prime)}
 = \sigma^2 \frac{w(\theta)}{\int 2\pi \theta^\prime\,
  \mathrm{d}\theta^\prime\, w^2(\theta^\prime)}
\end{equation}
Note that the kernel is unchanged if we rescale the fluctuation field,
which rescales $w$ and $\sigma^2$ by identical factors.  Thus the
kernel choice depends only on the shape, not the amplitude, of
$w(\theta).$ 

Our density estimator at field point $x$ thus becomes
\begin{equation}
\label{kde3}
\hat\delta(x) = \sigma^2 \frac{ \frac{1}{n_T} \sum_T w(\theta_{xT}) - 
 \frac{1}{n_R} \sum_R w(\theta_{xR})}{\frac{1}{n_R} \sum_R
 w^2(\theta_{xR})}.
\end{equation}

We caution the reader that the normalization in the denominator of equation~(\ref{kde3}) yields an unbiased estimator only when the kernel is chosen to match the true $w(\theta)$ for all $\theta.$  There are also practical issues to address with this KDE when the tracers are sparse and the value of $\hat\delta$ is not small, because the $(1+b\hat\delta)$ terms in the likelihood can yield non-physical (and non-mathematical!) probabilities $\le 0.$ These and other complications of the application of this method to real data will be explored in future work.

\subsection{Missing tracers}
\label{sec:missingtracers}
One practical difficulty with clustering-$z$ estimation has been that
tracers are not available over the full redshift range of the targets,
which leaves $n(z)$ indeterminate in these redshift gaps, and makes it
impossible to apply a normalization condition to the derived $n(z)$.

In the HBM method herein, a region of redshift, or of the sky, where we
have no tracers is equivalent to having to marginalize \eqq{pfz2} over
the unknown values of $\delta_z(\theta).$  For a single source galaxy,
this marginalization is equivalent to setting $\delta=0$, since
$\langle \delta \rangle = 0$ by construction.  We therefore can
proceed by simply setting $\delta_{iz}=0$ in \eqq{pzf3} for the
posterior within any angular or redshift gaps in the tracer coverage.
When there are multiple galaxies in such gaps, marginalizing
(\ref{pfz2}) over \fields\ is not necessarily equivalent to setting
$\fields=0$ because of the influence of the prior
$p(\fields,\pi_\delta)$ on correlations between different
$\delta_{iz}$.  But we are already ignoring such effects by using the
estimator $\hat\delta$ in place of marginalization over the $\delta$
process.

In short, where we do not have tracer information, we do not try to
use it. In these regions, the posterior probability simply reverts to
the photometric-only form.  Thus while our HBM formalism does not
eliminate degeneracies in pure clustering-$z$ inference from gaps in the
tracers, it does admit a combination with \photoz\ in a principled
fashion in which the photometric data resolves the clustering
degeneracies to the extent it can.  A similar statement can be made
about the bias-density degeneracy inherent to clustering-$z$
measurements; the HBM does not eliminate this degeneracy, but does
break this degeneracy to the extent possible by combination of the two
methods. 

\subsection{Relation to standard clustering redshifts}
The methods of \citet{Newman2008,Menard2013,Schmidt2013} are all predicated on the
assumption that, when both data and tracers are restricted to a redshift
bin $z$,  the cross-correlation $w_{DT}(\theta)$ between the data
(targets) and the tracers is related to the
tracer auto-correlation by
\begin{equation}
w_{DT}(\theta) = b_z w_z(\theta).
\end{equation}
[In this subsection there is no division of the target sample into phenotypes.]
If this is true, and we consider the density fields in disjoint bins
to be uncorrelated, then the cross-correlation between the full data
sample $D$ and the targets $T\!z$ in bin $z$ becomes
\begin{align}
w_{DT\!z}(\theta) & = f_z b_z w_z(\theta) \\
\label{wdtz1}
\Rightarrow \qquad f_z b_z & = \frac{w_{DT\!z}(\theta)}{w_z(\theta)}
\end{align}
Each angular bin $\theta$ can thus provide an estimator for the
quantities $q_z\equiv f_z b_z.$  If we define $XY_\theta$ as the pair
counts between population $X$ and $Y$ in a bin $\theta$ of angular
separation, divided by the densities $n_X$ and $n_Y$, then a standard
estimator using \eqq{wdtz1} is
\begin{equation}
\hat q_{z\theta} = \frac{DT\!z_\theta -
  DR_\theta}{w_z(\theta)\,DR_\theta}.
\end{equation}
using the populations data galaxies ($D$), randoms ($R$), and tracers
at redshift $z$ ($T\!z$).

All three of the above papers effectively produce a single
estimator for $q_z$ through a weighted sum over the $\hat q_{z\theta}$.  
[In these papers, the estimation of $q_z$ is sometimes described as doing a least-squares fit of a model to the observed $w(\theta)$ data.  When the free model parameter is the overall clustering amplitude, the least-squares solution can be expressed as an equivalent weighted sum.]
An optimized estimator is easily
derived under the assumption that errors are dominated by shot noise
in the $DT\!z$ pair counts. The weights are proportional to $w_z(\theta)\cdot
DR_z$ and the estimator can be rewritten as
\begin{align}
\hat q_z & = \frac{ \sum_\theta w(\theta) \left(DT\!z_\theta -
           DR_\theta\right)}
{\sum_\theta w^2(\theta) DR_\theta }\\
\label{wzest}
 & = \frac{ \frac{1}{n_{T\!z}} \sum_{i\in D, j \in T\!z} w(\theta_{ij})
 - \frac{1}{n_{R}} \sum_{i\in D, j \in R} w(\theta_{ij})}{
\frac{1}{n_{R}} \sum_{i\in D, j \in R} w^2(\theta_{ij})}.
\end{align}

It is clear that clustering methods alone can never break the
degeneracy between redshift distributions $f_z$ and bias $b_z$.  The similarity between \eqq{wzest} from the clustering-$z$ methods and the optimal KDE estimator for $\hat\delta_z$ in \eqq{kde3} suggests that these quantities are related.  In Appendix~\ref{sec:equivalence}, we demonstrate that in the limit where the \photoz's are uninformative, and the clustering is in the Poisson limit, that maximization of the posterior using \eqq{pzf3} and the optimal KDE in \eqq{kde3} yields the same result as the standard 2-point methods.  This strongly suggests that our joint constraint method makes use of all the information that is used by the standard clustering-$z$ methods.  Both methods are subject to the same degeneracy between $b_z$ and $f_z$, though our method allows the \photoz\ information to potentially break this degeneracy.

\section{Sampling and Priors}
\label{sec:sampling}

Now we consider the problem of simultaneously constraining the redshift and type probability distributions of populations of galaxies and their individual constituents. It is difficult to sample all variables simultaneously from the joint posterior $p(\fractions, \redshifts, \types, \biases | \features, \positions)$ in \eqq{pfz2}. It is possible, however, to draw samples from this posterior using a three-step Gibbs sampler because the conditional posterior distributions can be sampled. In the simulated data used to illustrate the method in this paper, we will assume that the biases $b^{t}_{z} = 1$, \ie\ assume that the tracers and targets have the same fluctuations. Sampling of the \biases\ will be demonstrated in a future publication, and hence we will be describing a two-step sampler in this section.

Each iteration of the Gibbs sampler comprises two steps which are (i) drawing a sample of $\fractions$ from $p(\fractions|\redshifts,\types,\features,\positions)$ and (ii) drawing pairs of $z_i, t_i$ for each galaxy $i$ from $p(z_i,t_i|\fractions,F_i,\theta_i)$ using the newly drawn $\fractions$.  The conditional distributions can be read directly from the joint distribution in \eqq{pfz2}.  We will make the further simplification that the effective area $A^t$ is independent of type and therefore $A(\fractions)$ is constant. Next we detail the expressions used in each of the two steps:

\begin{enumerate}[label=(\roman*)]
	\item The conditional posterior on $\fractions$ depends on the counts of sources of $\redshifts$ and $\types$ (in the last iteration), with $\counts = \{\nzt\}$ where $\nzt$ is the number of sources assigned to redshift $z$ and phenotype $t$:
		\begin{equation}
			p(\fractions|\redshifts,\types,\features,\positions) \propto p(\fractions) \prod_{z,t} f_{zt}^{\nzt} .
		\end{equation}
It also depends on the prior information on $\fractions$, $p(\fractions)$. The prior condition that $\sum f_{zt}=1$, and $0 \leq f_{zt} \leq 1$, allows us to write the conditional posterior on \fractions\ as a Dirichlet distribution, as we will explore next in \S\ref{sec:prior}. This allows us to draw a realization of $\fractions$ that we will use in the next step of the Gibbs sampler.       
	\item For each galaxy, the posterior for the $z_i,t_i$ pair conditioned on \fractions\ is
		\begin{equation}
	p(z_i,t_i |\fractions,F_i,\theta_i) \propto \likeli_{it_i} \, f_{t_iz_i} \, \left(1+ \hat\delta_{iz_i}\right)
		\end{equation}
		where apart from using the $\fractions$ obtained in the first step of the sampler (i), we make use of the measurement likelihood and the clustering terms discussed above. The sampling in this step (ii) will produce pairs of $z,t$ for each galaxy that constitute the next realization of $\counts = \{\nzt\}$, to be used in the step (i) of the next iteration of the Gibbs sampler.   
\end{enumerate}

The two-step Gibbs sampler allows us to explore the joint posterior distribution and hence to get samples of the redshift distribution of the full galaxy population as well as the individual redshift probability distributions.  

\subsection{Introducing a prior}
\label{sec:prior}
\vspace{1mm}

Let us now revisit the conditional posterior distribution on step (i) of the Gibbs sampling, $p(\fractions|\redshifts,\types,\features,\positions)$. Such distribution depends only on the number counts of $\redshifts$ and $\types$ pairs for the sources, so we can write $p(\fractions|\redshifts,\types,\features,\positions) = p(\fractions|\counts)$. Now we can use Bayes' theorem to write the posterior on the distribution parameters $\fractions$ as a likelihood of the binned data given the $\fractions$ probabilities times a prior on those probabilities:
\begin{equation}
	p(\fractions|\counts )  \propto p(\counts|\fractions) p(\fractions)
\end{equation}
The likelihood of the binned counts \counts\ given the probabilities $\fractions$ follows a multinomial distribution: 
\begin{equation}
	p(\counts |\fractions) = N! \prod_{z=1}^{N_z}\prod_{t=1}^{N_t}\frac{f_{zt}^{n_{zt}}}{n_{zt}!}
	\label{multinomial}
\end{equation}
The Dirichlet distribution is the conjugate prior of the multinomial distribution. That has the great advantage that if we choose the prior to follow a Dirichlet distribution, which appropriately fulfills $0 \leq f_{zt} \leq 1$ and $\sum_{zt} f_{zt} = 1$, then the posterior distribution will also be Dirichlet distributed. 

In getting the conditional posterior distribution of interest, $p(\fractions|\counts)$, we will distinguish between two cases, uninformative and informative priors.  
\subsubsection{Uninformative prior}
The Dirichlet distribution corresponding to an uninformative, uniform prior is:
\begin{equation}
p(\fractions) = (N_zN_t -1)!\, \delta_D \left(1-\sum_{zt}f_{zt}\right)\prod_{z=1}^{N_z}\prod_{t=1}^{N_t}\Theta(f_{zt}).
\end{equation}
Combining with the likelihood in \eqq{multinomial}, we get the following posterior: 
\begin{align}
	\nonumber
	p(\fractions|\counts) & = (N + N_zN_t -1)!\, \delta_D\left(1-\sum_{zt}f_{zt})\right) \\
 & \phantom{=} \times  \prod_{z=1}^{N_z}\prod_{t=1}^{N_t}\frac{\Theta(f_{zt})f_{zt}^{n_{zt}}}{n_{zt}!} 
\nonumber \\
	& \equiv \mathrm{Dir}(\counts)
	\label{uninformative}
\end{align}
This is a Dirichlet distribution parameterized by the counts \nzt~from the last iteration of the sampler. For brevity, we will write all subsequent Dirichlet distributions using the compact form of \eqq{uninformative}.  
\subsubsection{Informative prior}
Alternatively, an informative prior on the coefficients $\fractions$ may come from a representative (random) subset of galaxies with known $z,t$---\eg\ from a complete spectroscopic survey of a random subsample of targets.  If $\priorcounts=\{\mzt\}$ are the counts of this prior sample found at each $z,t$ pair, then the prior distribution of $\fractions$ follows a Dirichlet distribution with parameters \priorcounts, and hence the conditional posterior follows a Dirichlet on the data counts from the last iteration plus the prior counts:
\begin{equation}
p(\fractions|\counts) \sim \mathrm{Dir}(\counts + \priorcounts).
\end{equation}
In this way, it is very clear how the number of galaxies in the prior, $\mzt$, which will be small relative to the number of galaxies in the full sample, $\nzt$, will determine the relative importance of the prior in the HBM. The effect of the prior will be explored further in the next section, with a demonstration on simulations.   

\section{Demonstration on simulations}
\label{sec:sims}

We now present a simple simulation of galaxy survey data and test our methodology by exploring some of its main features. We adopt simple galaxy properties and noise distributions as the method can be easily adapted to account for realistic effects. 

In the simulation, we choose to work in a space where $z \in  [0,1]$ with 50 equally spaced bins, and we have 50 different galaxy types, which for convenience will be defined between 0 and 1 like the redshifts, so that $t \in [0,1]$ and $N_z = N_t = 50$. For each redshift bin (or slice) we generate $\delta_z$ from a Gaussian Random Field (GRF) with a resolution of 1024x1024 pixels that we will use to define galaxy positions in a way to simulate galaxy clustering. For every redshift slice, the fields are drawn from a power spectrum $P(k) \sim k^{-3}$, where $k$ is a given scale in Fourier space, and they are chosen to be uncorrelated between different redshifts. The RMS value of the GRF is around 2.5 for each slice, and  hence we clip the GRF to maintain non-negative density.
If $\rho(x,y,z) = \rho(\theta,z)$ is the value of the field at each line of sight and redshift, we define the overdensity field as:
\begin{equation}
	\delta(\theta,z) = \frac{\rho(\theta,z)}{\bar{\rho}(z)} - 1, 
	\label{eq:delta}
\end{equation}
where the mean field $\bar{\rho}(z)$ is taken, for each redshift slice, over all the 1024x1024 pixels in it. In reality, a (biased) estimate of this field can be obtained from a population of tracers with well-known redshifts, as described in Section \ref{sec:density}. For this demonstration, we will take the field $\delta(\theta,z)$ as known, rather than create a tracer population and KDE estimator.  This demonstration will also assume known bias ($b=1$) and perfect detection/selection ($p(s)=1$).

\begin{figure}
\includegraphics[width=\columnwidth]{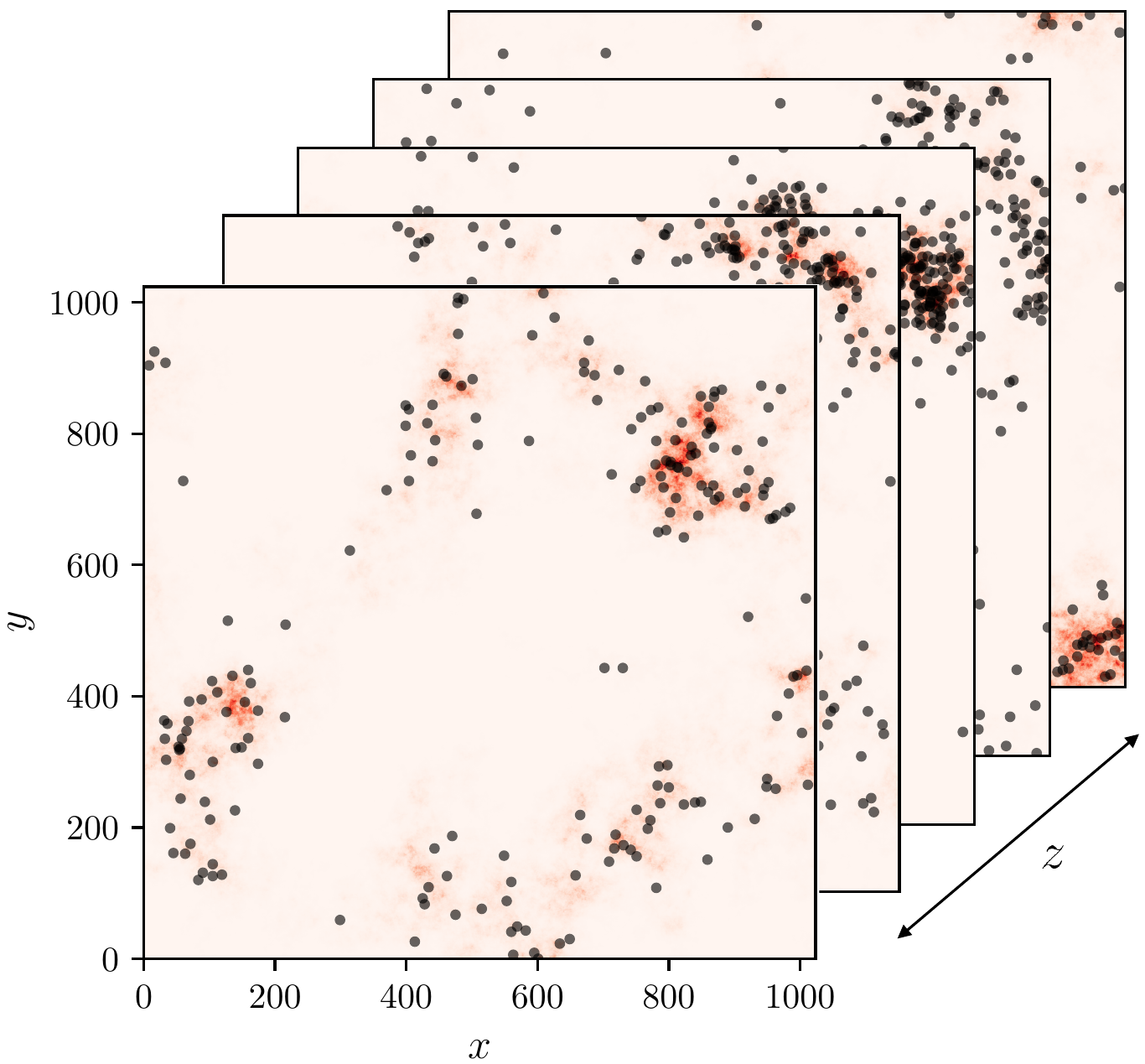}
	\caption[]{A graphical representation of the clustering present in the simulation used in this work. The plot shows five random redshift slices of the simulation, each of them showing the Gaussian Random Field (GRF) $\delta$, in red, and a subset of the galaxies placed in the slice (a 5\% random subset). As explained in the main text, there are no correlations between  $\delta$ fields of distinct slices. The axis in the plot correspond to the 1024x1024 resolution of the GRFs.}
\label{}
\end{figure}

\begin{figure*}%
    \centering
\subfloat{
    \includegraphics[width=0.48\textwidth]{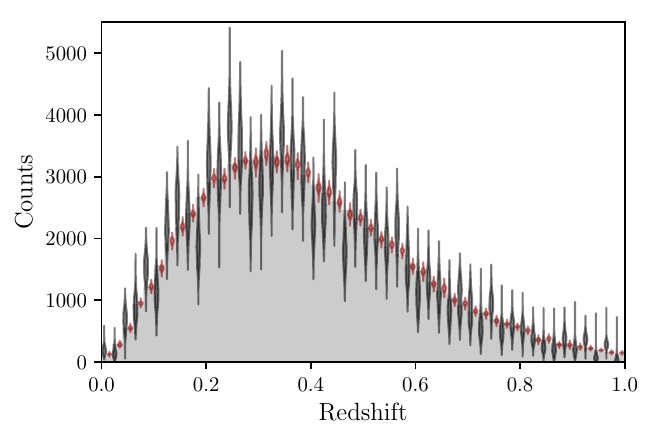} 
} 
\hfill
\subfloat{
    \includegraphics[width=0.48\textwidth]{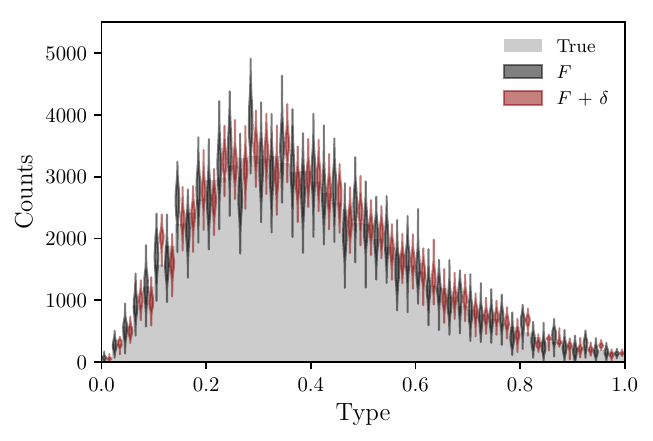} 
}
	\caption{(\textit{Left panel}): Redshift posterior distribution, marginalized over type, for the HBM method with and without the inclusion of clustering in the posterior. (\textit{Right panel}): Posterior distribution of galaxy type, marginalized over redshift, for the same cases. In both plots the true distributions of redshift and type, respectively, are shown for comparison with the recovered distributions. Note the strong effect of the clustering addition in recovering a tighter redshift distribution. }%
    \label{fig:violins}%
\end{figure*}

\begin{figure}
\includegraphics[width=\columnwidth]{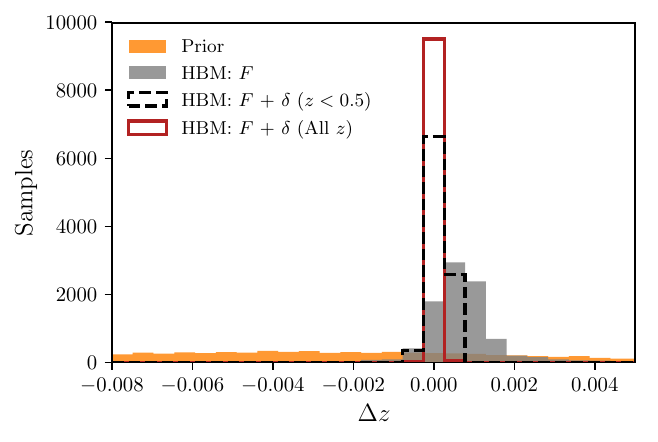}
\caption[]{Distribution of the $\Delta z$ metric, the difference between the mean of the recovered and true redshift distributions, for the redshift distribution posteriors displayed in Figures \ref{fig:violins} and \ref{fig:violinzczlim}. The effect of the clustering addition is clear again. The constraining power of the prior is also shown in this plot. }
\label{fig:dz_nobias}
\end{figure}

The simulation consists of $N = 10^5$ galaxies with types $\types$, redshifts $\redshifts$, features $\features$ and positions $\positions$ produced as follows:
\begin{itemize}
	\item We draw a galaxy type $t$ for each galaxy from the following distribution:
\begin{equation}
	\label{tdist}
	p(t) \propto t^a \exp^{-(t/t_0)^a},
\end{equation}
	with $a = 1.5$, $t_0 = 0.3$.  
	\item We assign redshifts to these galaxies depending on their type by using a simple model whereby a given galaxy type can only span, at most, three possible redshift values:
\begin{equation}
	\label{ztdist}
	p(z|t) = 
\left\{
  \begin{array}{ll}
	  \mbox{if } t=0\; & \left\{
      \begin{array}{ll}
                  0.8  & \mbox{ } z=t \\
                  0.2 & \mbox{ } z=t+0.02
          \end{array}
  \right.
   \\ \\
				   \mbox{if } t=1\;& \left\{
      \begin{array}{ll}
                  0.8  & \mbox{ } z=t \\
                  0.2 & \mbox{ } z=t-0.02
          \end{array}
  \right.
  \\ \\
	  \mbox{else }\;&
\left\{
      \begin{array}{ll}
                  0.6  & \mbox{ } z=t \\
                  0.2 & \mbox{ } z=t\pm0.02
          \end{array}
  \right.
 
          \end{array}
  \right.
\end{equation}
	\item We generate an observable feature $F$ for each galaxy using a very simplified model: Based on each galaxy's type $t$, we create a one-dimensional \textit{flux} such that the measurement likelihood is Gaussian with some variance $\sigma_F^{2}$, as:
	\begin{equation}
		p(F|t) = \likeli_{Ft} = \mathcal{N}(t-F,\,\sigma_F^{2}). 
	\end{equation}
	\item Finally, given each galaxy's redshift, we draw a position $\theta$ for each of them following:
	\begin{equation}
		p(\theta|z) \propto (1+\delta(\theta,z)), 
	\end{equation}
		where the field $\delta$ is taken exactly using the GRF as in \eqq{eq:delta}. 
\end{itemize}
Figure 2 presents a graphical description of the simulation described here, showing some example GRFs and galaxies drawn.

Next we will use this simulation to test the methodology described earlier in this paper. In applying the sampling methods described in Section \ref{sec:sampling}, we will obtain samples of the full posterior redshift distributions of galaxy populations and their individual members. These samples can be directly used in cosmological analyses to propagate the uncertainties coming from photometric redshift estimation. However, in analyzing the results, we will also find it useful to define a simple metric that allows us to easily compare between different cases or variations of the scheme. For that, we compute the difference in the mean between the estimated and the true redshift distributions, where the former we can compute for each sample $j$ of the redshift distribution coming from the HBM implementation (or the Dirichlet prior) and the latter is a fixed quantity: 
\begin{equation}
\Delta z_j = \left \langle z_{{\rm est},j} \right \rangle - \left \langle z_{\rm{true}} \right \rangle. 
\end{equation}
We will show both the distribution of these $\Delta z_j$ metric for the samples $j$ from the posterior, together with the median and standard deviation of such distributions.  

In all the tests performed in this Section, we run the chains of the Gibbs sampler using 4 different walkers, each with 2500 samples, for a total of $10^4$ samples per chain, and require the Gelman \& Rubin convergence metric \citep{Gelman1992} to be $R<1.03$ for considering the chain converged. The walkers are initialized at different realizations of the Dirichlet prior. All the chains run in less than an hour in a two-core laptop when using a Python implementation.   

\subsection{Fiducial results}

Now we present the results of a fiducial case in which we use a set of $8\times 10^4$ galaxies from our simulation as our photometric sample (we only know \features\ and \positions\ about these) and a random subset of $10^3$ galaxies as our prior (for these we know \redshifts\ and \types).    In this and other cases, we will use ``$F$'' to refer to analyses that use only photometric (feature) information, and ``$F+\delta$'' to refer to the combination of photometric information with clustering as per \eqq{pfz2}.

Figure \ref{fig:violins} shows the redshift and type distributions of the simulation together with the recovered distributions obtained with samples of the full posterior distributions, as violin plots. For the recovered distributions, both the $F$ only and the $F+\delta$ cases are shown. In the two cases, the redshift and type distributions of the photometric sample are effectively recovered, within the uncertainties displayed by the violins in each case. However, strong differences appear between the two cases in the magnitude of the uncertainties, especially for the redshift distribution. The observed features $\features$ in the photometric sample are used in the HBM scheme to tighten the posterior distributions compared to the prior, but these features $\features$ mainly inform the types $\types$ of galaxies, through the likelihood $\likeli_{Ft}$. In contrast, when adding clustering to HBM, that informs the redshift part of the problem, and can tighten the redshift distribution posterior significantly, as can be appreciated in the left panel of Figure \ref{fig:violins}.    

Figure \ref{fig:dz_nobias} shows the distribution of the $\Delta z$ metric computed from the redshift posteriors shown in the left panel of Figure \ref{fig:violins}. Again, the addition of clustering to the HBM method considerably sharpens the recovered $\Delta z$ distribution, without causing any apparent bias. Moreover, this plot includes the distribution of $\Delta z$ from samples of the Dirichlet prior. The median and standard deviation of the $\Delta z$ distributions shown in that Figure can be found in Table \ref{tab:fiducial}.    

\begin{figure}
\includegraphics[width=\columnwidth]{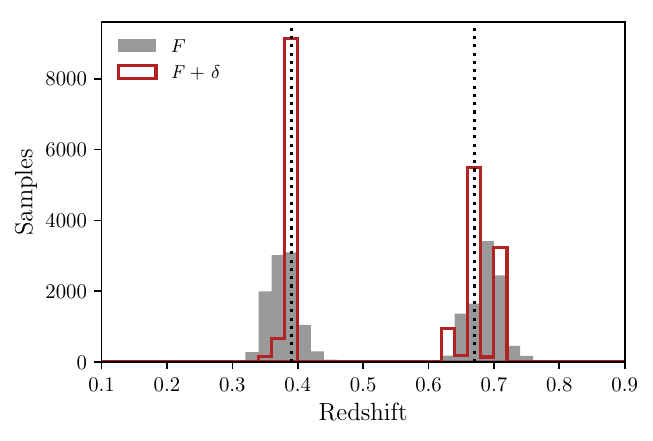}
	\caption[]{Posterior redshift probability distributions for two random galaxies in the sample, at redshifts $z=0.39$ and $z=0.67$ (dotted lines), using the HBM with and without clustering information. The addition of clustering can significantly sharpen the redshift posteriors of individual galaxies as well as that of the full population $n(z).$ }
\label{fig:gal_ex}
\end{figure}

So far we have looked at the sampling of the redshift and type distributions for the entire population. However, in the HBM approach we are sampling both the probability distributions of the populations together with those of their individual members. In particular, we are sampling the redshift probability distribution of each individual galaxy in the sample. Figure \ref{fig:gal_ex} shows the sampling of the redshift probability distributions of two random galaxies in the sample, for the HBM method with and without clustering information, showing the true redshift of the galaxies, for comparison. We can see how the addition of the clustering information sharpens the individual redshift probability distributions as well.

\subsubsection{Clustering only up to a given $z$}

\begin{figure}
\includegraphics[width=\columnwidth]{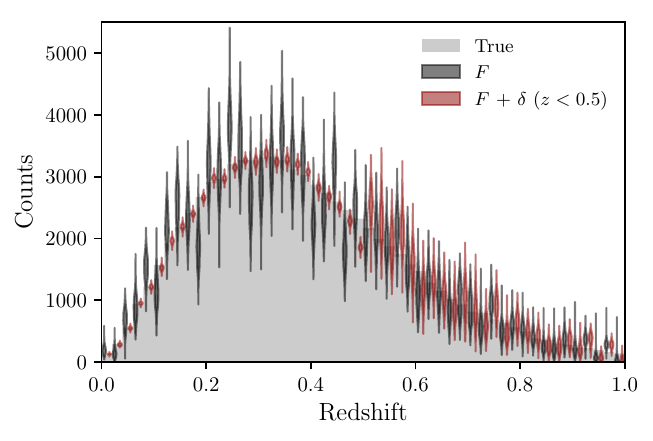}
\caption[]{Redshift posterior distribution, marginalized over type, for the HBM method and the addition of clustering to it. Here we can see that the method works properly when density information is available for only half of the redshift range ($z<0.5$), with the expected result that the posterior at redshifts with $z>0.5$ resembles the photometry-only case. }
\label{fig:violinzczlim}
\end{figure}

The framework for the addition of clustering information in this work enables the consistent usage of that information even if it is only available in a part of the redshift range of interest (see Section \ref{sec:missingtracers}). Here we explore this aspect of the method by using the exact knowledge of the $\delta$ field only in half the redshift range of the simulation. Figure \ref{fig:violinzczlim} shows the redshift distribution posteriors when using the knowledge of $\delta$ in the range $z<0.5$. The effect of clustering is apparent in the lower half of the redshift range, and the posteriors in the other half appear consistent with the HBM method without clustering information. More quantitatively, Figure \ref{fig:dz_nobias} and Table \ref{tab:fiducial} show how the $\Delta z$ metrics behave when using clustering information in only a part of the redshift range, demonstrating that it significantly helps constraining that metric over the HBM without clustering case, without being as constraining as using the clustering information in the entire redshift range.       

\begin{table}
	  \caption{Median and standard deviation of the $\Delta z$ distributions presented in Figure \ref{fig:dz_nobias}, showing the constraints for the prior and the HBM runs, with and without clustering information, for the fiducial case.  }
    \label{tab:fiducial}
    \begin{center}
    \begin{tabular}{lrr} 
		    \vspace{-1.5mm}
	    	    & Median ($\Delta z$) & Standard Deviation ($\Delta z$) \\
		    \vspace{-0.7mm}
	    \textit{Fiducial}	    &  &  \\
      \hline
	    Prior & -0.00356 & 0.00608 \\ 
	    HBM: $F$ & 0.00068 & 0.00235 \\ 
	    HBM: $F$ + $\delta$ ($z<0.5$) & 0.00012 & 0.00022 \\ 
	    HBM: $F$ + $\delta$ (All $z$) & 0.00003 & 0.00010 \\
    \end{tabular}
    \end{center}
\end{table}

\subsection{Biases in the prior}

So far in this Section we have shown how the method described in this work can provide an unbiased sampling of the redshift distribution of a galaxy population when we use a representative subsample of it as a prior. However, in reality, having a representative galaxy subsample with both spectroscopy and accurate photometry, so we know $\redshifts$ and $\types$ for those galaxies, can be very difficult. For this reason, we will study how the method is sensitive to biases in the subsample used for the prior and, especially, if the method is able to overcome such biases for the estimated distributions. In the analysis of the possible biases that can plague our prior sample, we will split the prior expression as in \eqq{pzt}, $p(z,t) =p(z | t) p(t) = f^t_z f_t$, and we will explore the effects of biasing the prior separately in $f_t$ and $f^t_z$, and then in both simultaneously.

\subsubsection{Bias in $f_t$}

We first introduce a bias in the $f_t = p(t)$ part of the prior. The $p(t)$ distribution gives the abundance of galaxies in the different regions of the observed feature space, for instance it would correspond to the density of galaxies in each cell of a self-organized map of galaxy colors. In presenting the details of the simulation, we showed that a type $t$ for each galaxy is drawn from the $p(t)$ distribution in \eqq{tdist}, with $a = 1.5$, $t_0 = 0.3$.  Now, in order to select galaxies for the prior, we draw a subset of 1000 galaxies from the simulation following the distribution in \eqq{tdist}, with $a=1.3$ and $t_0 = 0.25$, while keeping the $f^t_z$ relation as in \eqq{ztdist}. Because of the relation between $\redshifts$ and $\types$, this bias in $f_t$ produces a bias in redshift, of the order of $\Delta z \lesssim -0.02$.   

The $\Delta z$ constraints for this case coming from the prior and the HBM method with and without clustering information are shown in the upper panel of Figure \ref{fig:dz_biased} and in Table \ref{tab:biased}. In there, we can clearly see the bias in the prior, and we can also see how the HBM method, both with and without the usage of clustering information, is able to correct for that bias in the derived redshift distribution for the sample.  The $F$-only analysis reduces redshift bias by $20\times$ in this case, because the bias is due to a mis-estimate of the distribution of galaxy phenotypes, which can be corrected with photometry from the full sample of sources, even if the photometry is noisy. 

\begin{table}
	  \caption{Median and standard deviation of the $\Delta z$ distributions presented in Figure \ref{fig:dz_biased}, showing the constraints for the prior and the HBM runs, with and without clustering information, for the cases with biases in the prior.  }
    \label{tab:biased}
    \begin{center}
    \begin{tabular}{lrr} 
		    \vspace{-1.5mm}
	    	    & Median ($\Delta z$) & Standard Deviation ($\Delta z$) \\
		    \vspace{-0.7mm}
	    \textit{Bias in $f_t$}	    &  &  \\
      \hline
	    Prior & -0.02644 & 0.00614 \\ 
	    HBM: $F$ & -0.00132 & 0.00181 \\ 
	    HBM: $F$ + $\delta$ ($z<0.5$) & 0.00015 & 0.00024 \\ 
	    HBM: $F$ + $\delta$ (All $z$) & -0.00010 & 0.00009 \\
		    \vspace{-1.7mm}
       &  & \\
		    \vspace{-0.7mm}
	    \textit{Bias in $f_z^t$}	    &  &  \\
      \hline
      Prior & -0.00632 & 0.00628 \\ 
      HBM: $F$ & -0.00493 & 0.00338 \\ 
	    HBM: $F$ + $\delta$ ($z<0.5$) & -0.00061 & 0.00021 \\ 
	    HBM: $F$ + $\delta$ (All $z$) & -0.00024 & 0.00010 \\
		    \vspace{-1.7mm}
       &  & \\
		    \vspace{-0.7mm}
	    \textit{Bias in $f_t$ \& $f_z^t$}	    &  &  \\
      \hline
	    Prior & -0.03083 & 0.00623 \\ 
	    HBM: $F$ & -0.00496 & 0.00124 \\ 
	    HBM: $F$ + $\delta$ ($z<0.5$) & -0.00094 & 0.00020 \\ 
	    HBM: $F$ + $\delta$ (All $z$) & -0.00026 & 0.00010 \\
    \end{tabular}
    \end{center}
\end{table}

\subsubsection{Bias in $f^t_z$}

	We introduce a bias in the $f^t_z = p(z|t)$ part of the prior, which tells us about the relation between galaxy types (or observed features) and redshift, which is known as the \textit{color-redshift} relation if we are only using galaxy colors as features. We randomly draw galaxies from the simulation to be treated as ``spectroscopic,'' and for a spectroscopic source of type $t$, we assign a redshift with a biased modification of the relation in \eqq{ztdist}:
\begin{equation}
	\label{ztdist_biased}
	p(z|t) = 
\left\{
  \begin{array}{ll}
	  \mbox{if } t=0\; & \left\{
      \begin{array}{ll}
                  0.85  & \mbox{ } z=t \\
                  0.15 & \mbox{ } z=t+0.02
          \end{array}
  \right.
   \\ \\
				   \mbox{if } t=1\;& \left\{
      \begin{array}{ll}
                  0.75  & \mbox{ } z=t \\
                  0.25 & \mbox{ } z=t-0.02
          \end{array}
  \right.
  \\ \\
	  \mbox{else }\;&
\left\{
      \begin{array}{ll}
                  0.5  & \mbox{ } z=t-0.02 \\
                  0.4  & \mbox{ } z=t \\
                  0.1 & \mbox{ } z=t+0.02
          \end{array}
  \right.
 
          \end{array}
  \right.
\end{equation}
The bias in this relation produces a shift in the mean redshift of the spectroscopic prior, $\Delta z \sim -0.005$.  The $\Delta z$ constraints for this case coming from the sampling of the prior and the HBM method, with and without clustering information, are presented in the middle panel of Figure \ref{fig:dz_biased} and in Table \ref{tab:biased}. In this case, the HBM method without clustering information is unable to correct for the bias in the prior, as the observed features of the galaxies in the sample do not inform the $f^t_z$'s, only the $f_t$'s, which are unbiased in this case. Without the usage of clustering information, the color-redshift part of the problem is only informed by the prior. However, when using clustering in the HBM, we are adding direct information about redshift, and hence the method is able to significantly reduce the redshift biases in the prior.      

\subsubsection{Bias in $f_t$ and $f^t_z$}

\begin{figure}
    \centering
\subfloat{
    \centering
    \includegraphics[width=0.5\textwidth]{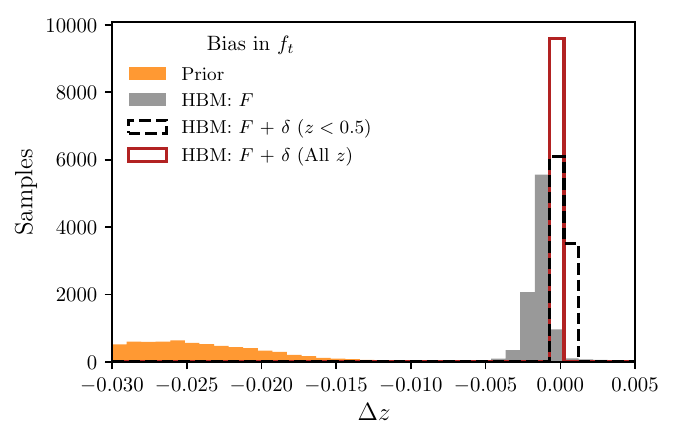} 
}

    \qquad
\subfloat{
    \centering
    \includegraphics[width=0.5\textwidth]{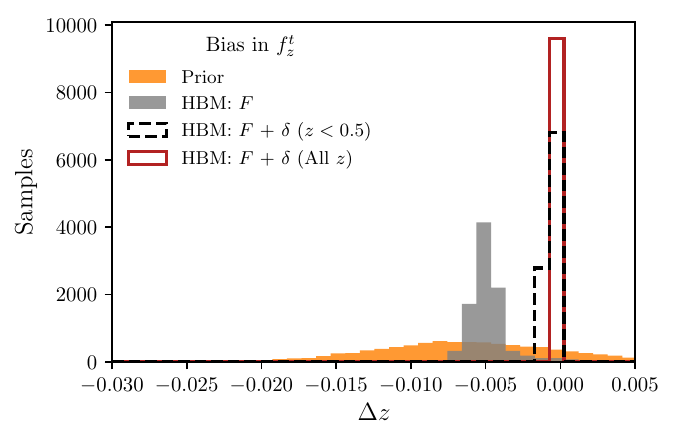} 
}

    \qquad
\subfloat{
    \centering
    \includegraphics[width=0.5\textwidth]{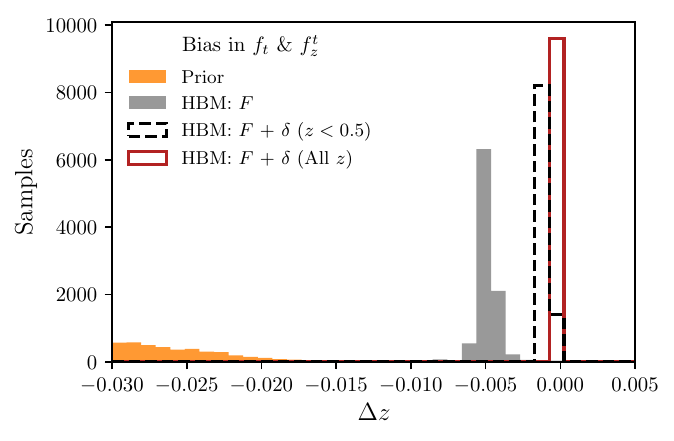} 
}

\caption[]{Distribution of the $\Delta z$ metric for the redshift distribution posteriors coming from the prior and the HBM methods, with and without using clustering information, in the presence of biases in the prior: bias in $f_t$ (upper), bias in $f_z^t$ (center), and bias in both $f_t$ and $f_z^t$ (lower). One can see how the HBM method without clustering information can correct biases in the $f_t$ part of the prior, but biases in the $f_z^t$ part can only be corrected by adding clustering information. }
	\label{fig:dz_biased}
\end{figure}

Now we combine the two types of biases analyzed before in a more realistic case where the prior is biased in both the type-redshift relation and the abundance of the different galaxy types, in the same way we introduced each of such biases above. The results are shown in the lower panel of Figure \ref{fig:dz_biased} and in Table \ref{tab:biased}. Again, the HBM method without clustering information is able to correct the bias coming from the prior having a different galaxy type distribution, but is unable to correct for a wrong type-redshift relation in the prior. On the other hand, the addition of clustering allows the HBM method to significantly reduce biases coming from the prior, as it is using the galaxy sample observed features to inform the galaxy type distribution and the clustering information to inform the redshift part.

\vspace{-3mm}
\section{Conclusions}
\label{sec:conclusions}
There are three main sources of information for estimating the redshift probability distributions of galaxies and ensembles of galaxies in a wide-field survey. First, we have prior information, which comes from a subset of galaxies, typically spanning a smaller area in the sky, for which we have both reliable redshift estimates and precise photometry. Then, using all the galaxies in the wide-field sample, there exist methods for estimating those distributions from either the photometry of such galaxies or the clustering of them against a tracer population with precise and well-characterized redshift estimates, which overlaps with the wide-field sample in the sky. However, one encounters important complications when trying to combine the independent answers from these last two approaches, since, for instance, the tracer population used in clustering methods typically only covers a subset of the redshift range of interest, and hence so do the derived redshift distributions. Also, it is very important to correctly propagate the uncertainties due to the limited size of the prior sample compared to the full population into the inference.     

The method presented in this paper is the first to combine these three sources of information in a unified and consistent scheme, using all the information available. The method consists of a hierarchical Bayesian model which allows us to sample both the redshift distribution of a galaxy sample and the individual redshifts of galaxies therein. It can be seen as a generalization of the approach in \citet{Leistedt2016} to include clustering information as well as an informative prior.  In the case where one marginalizes over the latent density fields \fields, the method gives a rigorous posterior probability for the redshift distribution $n(z)$, as well as individual $z_i$'s, if the galaxies are a Poisson sampling of a density field with known statistical properties.  The approximation that we make here to enable a straightforward sampling process is that the stochastic density field can be perfectly estimated from 
a kernel density estimator using the positions of a tracer population.  In the limit of weak photometric information and weak clustering, the maximum-posterior $n(z)$ estimator is shown to be equivalent to standard clustering-$z$ methods using 2-point correlation functions.  With the KDE approximation, the posterior $n(z)$ from our HBM no longer fully samples the effects of shot noise in the tracer population; this will be addressed in future improvements.

This paper's simple simulation, which is given knowledge of the true density field, demonstrates the benefits of adding clustering information into the photometric redshift estimation. This inclusion not only sharpens the posterior redshift probability distributions of galaxies and populations of galaxies but, more importantly, it can overcome biases in the prior which are not possible to resolve when only using galaxy photometry. That demonstrates the advantages of the combination of clustering and photometry into the same inference, where the usage of the two techniques separately would yield different answers which would be hard to combine or result in a failed validation cross-check. We also show how clustering information can be naturally used even if it is only available for a subset of the redshift range of interest, without producing any additional biases. 

The methodology described herein does not resolve redshift degeneracies that are intrinsic to the photometric and clustering methods: for \photoz's, there remains the possibility that the map of colors to redshifts is non-unique, \ie\ galaxies at distinct $z$ can have identical features $F$.  For clustering, it remains true that the redshift dependence of bias $b(z)$ is degenerate with the redshift distribution.  Our method does, however, exploit any cases in which one method might resolve the degeneracies of the other---for instance if the photometric features of some galaxy type $t$ restrict $p(z|t)$ to a small enough range of $z$ that variation in $b^t(z)$ can be assumed to be fairly small.  The success of this method in producing $n(z)$ to the accuracy desired for large cosmological surveys will therefore still depend on the details of the survey and the galaxy population; but we now have a means to harness all of the available information.

One of the attractive aspects of the hierarchical Bayesian model for photometric redshift estimation is that the sampling of the posterior distribution of $n(z)$ for the galaxy population allows us to correctly propagate the uncertainties in the estimation of that distribution into cosmological analyses of galaxy clustering and weak gravitational lensing. The model can also be extended to correct for other observational effects, such as the calibration of photometric zero points \citep{Leistedt2018}, and can trivially accommodate more complicated noise likelihood functions. Finally, the method presented here can eventually be incorporated as part of a fully Bayesian analysis of galaxy surveys, which would use all the available information to reconstruct the matter density field \citep{Jasche2010,Jasche2013}.

\section*{Acknowledgements}

The authors thank \`Alex Alarc\'on, Romain Buchs, Chris Davis, Daniel Gruen, Boris
Leistedt and Justin Myles for helpful conversations about this topic.  This work was
supported by grants AST-1311924 and AST-1615555 from the US National
Science Foundation, and DE-SC0007901 from the US Department of Energy.



\bibliographystyle{mnras}
\bibliography{/Users/csanchez/Dropbox/bibtex/library}



\appendix

\section{Equivalence to correlation-function methods}
\label{sec:equivalence}

Consider the case when there is no photometric information available
on our detected galaxies, 100\% selection rate, and the window function is independent
of redshift so that $A(\fractions)$ becomes some constant value $A.$
\eqq{pzf3} for the posterior reduces to
\begin{equation}
p(\fractions, \redshifts, \biases | \positions)
\propto p(\fractions) \prod_i f_{z_i}(1 +
b_{z_i}\hat\delta_{iz_i}).
\end{equation}
If we marginalize over \redshifts\ we obtain
\begin{align}
p(\fractions, \biases | \positions) & \propto \sum_\redshifts
                                          p(\fractions, \redshifts, \biases | \positions) \\
 & \propto p(\fractions) \prod_i \left[ \sum_{z} f_z \left(1+b_z\hat\delta_{iz}\right) \right] \\
 & = p(\fractions) \prod_i  \left[ 1+ \sum_{z} f_z b_z\hat\delta_{iz}\right], \\
\hat\delta_{iz} & \equiv \hat\delta_z(\theta_i).
 \end{align}
A minimal prior is that the redshift distribution fractions $f_z$ must be non-negative and sum to unity.  In this case we can effectively drop $p(\fractions)$ from this equation, and we find that the posterior probability depends only upon the quantities $q_z \equiv f_zb_z,$ \ie\ we recover the degeneracy between bias and $n(z)$ that is well known for pure clustering redshifts.  The log of the posterior is now
\begin{equation}
\log P(\fractions,\biases | \positions) = \sum_i \log \left[ 1 + \sum_z q_z \hat\delta_{iz}\right] + \textrm{const}.
\label{post1}
\end{equation}
We seek the estimated
values $\hat q_z$ which maximize this quantity.  Writing the estimated
projected
density along $\theta_i$ as $y_i \equiv \sum_z \hat q_z \hat\delta_z(\theta_i),$
the condition for posterior maximization at $\{\hat q_z\}$ is the set
of simultaneous equations
\begin{align}
0 & = \frac{\partial}{\partial \hat q_z} \sum_i \log (1+y_i) \\
 & = \sum_i \frac{\hat \delta_{iz}}{1 + y_i}.
\end{align}
We examine the limit $|y_i|\ll1,$ when the projected mass fluctuations are weak.  Note that this does \emph{not} require the spatial fluctuations $\delta$ to be weak.  The solution becomes
\begin{align}
0 & = \sum_i \left[ \hat\delta_{iz} (1 - y_i + y_i^2 - \ldots) \right] 
\label{solve1} \\
  & \approx \sum_i \hat \delta_{iz} - \sum_{z^\prime} q_{z^\prime}
    \sum_i \hat\delta_{iz} \hat\delta_{iz^\prime}
 + \sum_{z^\prime,z^{\prime\prime}} q_{z^\prime}q_{z^{\prime\prime}}
    \sum_i \hat\delta_{iz} \hat\delta_{iz^\prime} \hat\delta_{iz^{\prime\prime}}
\label{solve2} \\
  & \approx S_{1z} - q_z S_{2z} + q_z^2 S_{3z},
\label{solve3}
\end{align}
where we have defined
\begin{equation}
S_{nz} = \sum_i \hat\delta^n_{iz}.
\end{equation}
In going from (\ref{solve2}) to (\ref{solve3}) we have simplified a
matrix equation to a series of linear equations by assuming that the
independence of the density fields in distinct bins drives 
terms like $\sum_i \hat\delta_{iz} \hat\delta_{iz^\prime}$ to zero unless $z=z^\prime.$

If we make the approximation that the density fluctuations $\hat\delta$ are weak or symmetric in the sense that $S_{3z} S_{1z} \ll S^2_{2z}$, then the posterior is maximized at values $\hat q_z$ solving \eqq{solve3}:
\begin{equation}
\label{solve4}
\hat q_z  \approx \frac{S_{1z}}{S_{2z}} \left( 1 + \frac{S_{3z} S_{1z}}{S_{2z}^2} \right).
\end{equation}

At this point is it useful to calculate the expectation value of $S_{nz}$ under Poisson sampling of the density fields $\hat\delta_z:$
\begin{align}
\langle S_{nz} \rangle & = \int \mathrm{d}^2\theta \left\{ n \sum_{z^\prime} f_{z^{\prime}} \left[1 + b_{z^\prime} \hat\delta_{z^\prime}(\theta) \right] \hat\delta_z(\theta)^n \right\} \\
 & = nA \left( \mu_{nz} + f_z \mu_{(n+1)z} \right),
\end{align}
where $n$ is the mean source density, $A$ is the survey area, and we use the central moments of the density estimators
\begin{equation}
\mu_{nz} \equiv \left\langle \hat\delta^n_z(\theta) \right\rangle_\theta.
\end{equation}
We have $\mu_{1z}=0$ and $\mu_{2z}=\hat\sigma^2_z,$ the variance of the density estimator.

At this point we will make an approximation that the clustering is weak, $\sigma_z\ll<1,$ and that the central moments of the field satisfy $\mu_{nz} \lesssim \sigma^n_z,$ so that we can retain only leading terms in $\delta.$  In this limit we can approximate the denominator in \eqq{solve4} by its expectation value, $\langle S_{2z} \rangle = nA(\sigma^2_z + f_z \mu_{3z}) \approx nA \sigma^2_z.$ It can be shown that the correction term in parentheses in \eqq{solve4} will cancel the influence of the $\mu_{3z}$ term to first order.  We therefore arrive at a very simplified estimate of the maximum posterior solution $\hat q_z$:
\begin{align}
\hat q_z & \approx \frac{1}{\sigma^2_z} \sum_i \hat\delta_{iz} \\
           & = \frac{1}{N_D} \sum_{i\in D} 
\frac{ \frac{1}{n_{T\!z}} \sum_{j \in T\!z} w_z(\theta_{ij}) - 
 \frac{1}{n_R} \sum_{j \in R} w_z(\theta_{iR})}{\frac{1}{n_R} \sum_{j \in R}
 w_z^2(\theta_{ij})} \label{solve5} \\
 & \approx 
\frac{ \frac{1}{n_{T\!z}} \sum_{i\in D, j\in T\!z} w_z(\theta_{ij}) - 
 \frac{1}{n_R} \sum_{i \in D, j \in R} w_z(\theta_{iR})}{\frac{1}{n_R} \sum_{i \in D, j \in R}
 w_z^2(\theta_{ij})}. \label{solve6}
\end{align}
\eqq{solve5} makes use of the KDE for $\hat\delta$ from \eqq{kde3}.  In the final line we approximated that the denominator of (\ref{kde3}) is nearly constant for all galaxies $i$ in the data sample $D.$  The sets $D, T\!z,$ and $R$ are the target galaxies (data), the tracer population in redshift shell $z$, and a randomly distributed sample, respectively.  The resultant maximum-posterior estimate for $q_z$ is seen to be the same, under these weak-field approximations, as the Poisson-limited optimal estimator for $q_z$ from using two-point functions, \eqq{wzest}.

This demonstrates that in the weak-field, no-photometry limit, our posterior probability has a maximum at the same value for $q=fb$ as the standard correlation-function-based clustering-$z$ methodology.  We have not demonstrated, however, that the variances of the two estimators coincide.
Nonetheless the equivalence of the maximum-posterior estimator to the
standard 2-point estimator demonstrates that the latter is a limiting
case of our method, and we can expect our principled approach to the
$n(z)$ posterior to incorporate all the information of the 2-point
functions---especially if one can truly marginalize over the density
fields.


\bsp	
\label{lastpage}
\end{document}